%% file: pks0745.tex
\documentclass[useAMS,usenatbib]{mn2e}
\usepackage{amssymb,amsmath,epsfig,times}
\input{defs}

\title[PKS 0745-191 at the virial radius]{X-ray observations
  of the galaxy cluster PKS 0745-191: To the virial radius, and beyond}
\author[M. R. George et al.]{M. R. George,$^1$\thanks{Email: 
    mgeorge@astro.berkeley.edu  - Current address: Department of Astronomy,
    University of California, Berkeley, CA 94720, USA} A. C. Fabian,$^1$ J. S. Sanders,$^1$
  A. J. Young$^2$ and H. R. Russell$^1$\\
  $^1$Institute of Astronomy, Madingley Road, Cambridge CB3 0HA \\
  $^2$H. H. Wills Physics Laboratory, University of Bristol, Tyndall Avenue, Bristol BS8 1TL \\
}
\date{Accepted 22 January 2009}

\begin{document}

\maketitle

\begin{abstract}
We measure X-ray emission from the outskirts of the cluster of
galaxies PKS 0745-191 with \emph{Suzaku}, determining radial profiles of
density, temperature, entropy, gas fraction, and  mass. These
measurements extend beyond the virial radius for the first time,
providing new information about cluster assembly and the diffuse 
intracluster medium out to $\sim1.5
r_{200}\,(r_{200}\simeq1.7\mpc\simeq15')$. The temperature is found to
decrease by roughly 70 per cent from $0.3-1r_{200}$. We also see a
flattening of the entropy profile near the virial radius and consider
the implications this has for the assumption of hydrostatic
equilibrium when deriving mass estimates. We place these observations
in the context of simulations and analytical models to develop a
better understanding of non-gravitational physics in the outskirts of
the cluster.
\end{abstract}


\section{Introduction}
The outskirts of galaxy clusters present an opportunity to study the
formation of large scale structure as it happens. Beyond the core
where non-gravitational processes such as cooling flows and feedback
from active galactic nuclei can dominate activity, clusters are
expected to be more relaxed and to follow self-similar scaling
relations \citep{kaiser86}. The virial radius can be thought of as a border
between regions of equilibration and infall, so merger activity from
accreting material may play an important role in the outer dynamics.

As the largest virialized systems in the universe, clusters of
galaxies can be useful in constraining cosmological models. The
position of clusters at the high end of the mass spectrum and their
evolution from the initial density perturbations makes
them sensitive probes of the scale of these fluctuations, $\sigma_8$,
and the density of matter in the universe, $\Omega_{\rm m}$
\citep{white93,eke96}. Additionally, the evolution with redshift of
cluster properties such as the gas mass fraction can be used to study other
portions of the cosmic energy density budget \citep[\emph{e.g.,}
][]{allen08}. For a review of the cosmological importance of galaxy clusters,
see \citet{voit05}. 

To understand cluster properties, it is important that the relationships
between observables and derived quantities are well-calibrated. A
common example is that temperature and density profiles can be
measured from X-ray spectra and used to determine a cluster's mass
under the assumptions of spherical symmetry and hydrostatic
equililbrium between the gas pressure and gravitational
potential. Though numerical simulations predict a decline in
temperature towards the virial radius \citep[\emph{e.g.,}
][]{evrard96,frenk99, loken02}, observations at smaller radii have produced
inconsistent results. Some analyses have found declining profiles
\citep[\emph{e.g.,} ][]{markevitch98,degrandi02,vikhlinin05,pratt07},
while others have seen a scattering of slopes consistent with flat or
even increasing profiles \citep[\emph{e.g.,}
][]{irwin99,zhang04,arnaud05}. These studies typically do not measure
the temperature profiles much further out than half the virial radius,
even with \emph{Chandra} and \emph{XMM}. Different methods of
estimating the virial radius, using either scaling relations or direct
determination from the mass profile, make comparisons difficult, but
very few measurements have been made of cluster gas temperatures out
to the virial radius \citep[\emph{e.g.,} ][]{solovyeva07,
  reiprich08}, and to our knowledge none have been reported beyond it.

Until recently, X-ray detectors have been unable to measure the temperature
of the intracluster medium (ICM) out to the virial radius due to its
low surface brightness in the outskirts relative to background
noise. Typically a mass profile such as that of \citet*[][
NFW]{navarro97} is used to extrapolate mass estimates to distances
greater than those observed. One aim of our work is to measure
properties of the ICM out to the virial radius in order to check the
reliability of such extrapolations. For our purposes, we equate the
virial radius with $r_{200}$, the radius within which the mean total
density is 200 times the critical density of the universe at the
redshift of the cluster.

The low orbit of \emph{Suzaku} places it within Earth's magnetopause,
giving it a significantly lower and more stable particle background
compared to \emph{Chandra} and \emph{XMM-Newton}. \citet{reiprich08}
have recently leveraged this ability for low surface brightness
observations in the outskirts of A2204, measuring the temperature
nearly to $r_{200}$. Other cluster observations with
\emph{Suzaku} have demonstrated its capacity for temperature and
abundance measurements \citep[\emph{e.g.,} ][]{sato07}, providing an
outline for some of the methods used here.

In this work, we present \emph{Suzaku} observations of PKS 0745-191,
a relaxed, cool core cluster that is the brightest in X-rays beyond $z=0.1$
\citep{fabian85,edge90,allen96}. Previous measurements have found a
mean gas temperature in the range of $6.4-8.5\kev$, depending on the
models used and regions studied \citep*[][ and references
therein]{chen03}. \citet{allen96} find good agreement between the
central mass from X-rays and that determined from a strongly lensed
arc, but the results of \citet{chen03} disagree, finding a factor of 2
smaller mass from \emph{XMM} data \citep[see also ][ for a reanalysis
of the \emph{XMM} data]{snowden08}. We describe our observations in the
next section, followed by details of the spectral analysis in
Section~\ref{analysis}, and the resulting profiles in
Section~\ref{results}. In Section~\ref{discussion} we place our
findings in the context of other cluster studies, and we conclude the
paper in Section~\ref{summary}. For distance scales, we use the
cosmological parameters $H_0=70\kmps\mpc^{-1}$ and
$\Omega_{\rm m}=1-\Omega_{\Lambda}=0.3$, giving an angular scale of
$113\kpc$ per arcminute at the cluster's redshift, $z=0.1028$. The
plotted and quoted error ranges are $1\sigma$ statistical
uncertainties except where otherwise stated.

\section{Observations and Data Reduction}

\begin{figure}
  \begin{center}
    \leavevmode
      \epsfxsize=8cm
      \epsfbox{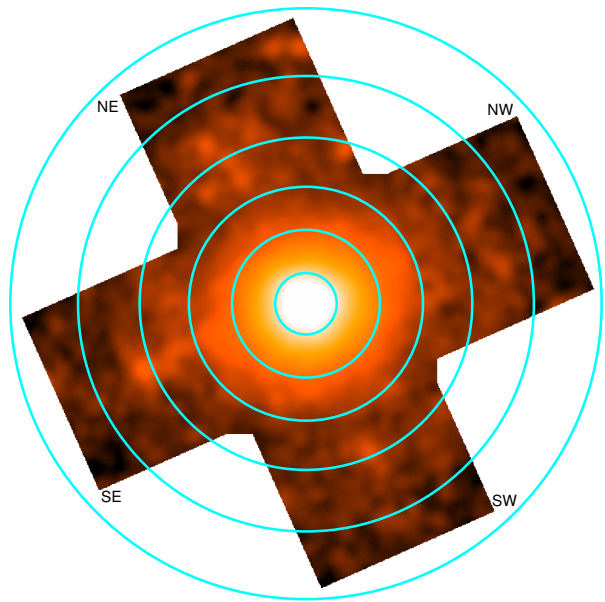}
      \caption{Exposure-corrected mosaic of PKS 0745-191, smoothed
        with a Gaussian of radius 1'. Background
        subtraction and vignetting correction have been omitted, and
        calibration regions and point sources have not yet been
        excised. Ring radii are 2.5', 6', 9.5', 13.5', 18.5', and 24'.}
      \label{image}
  \end{center}
\end{figure}

\emph{Suzaku} observations of PKS 0745-191 were taken between 2007 May 11-14
in five separate fields of roughly $32\ks$ each. Details of these
pointings are listed in Table~\ref{obsdetails}. The central pointing
is aimed toward the peak of the cluster emission and the others are
positioned adjacently, with ~4' overlap at each chip edge, as in
Fig.~\ref{image}. We use only data from the X-ray Imaging
Spectrometer \citep[XIS:][]{koyama07}. We exclude the back-illuminated
detector, XIS1, despite its higher sensitivity at low energies, because it
also has a significantly higher particle background level than the two
front-illuminated sensors, XIS0 and XIS3, which have similar
responses. There is an offset between the measured temperatures of the
BI and FI chips, and we discuss this contribution to our systematic
uncertainties in Section~\ref{uncertainties}. We simultaneously
analyze data from the two FI sensors to 
effectively double the exposure time while keeping the noise
low. Observations were taken in normal clocking and editing modes 
with spaced-row charge injection on, and the data have been processed with
the energy scale calibration of v2.1.6.16. The data preparation
described below was carried out using {\sc xselect} 2.4.1 and {\sc
  ftools} version 6.5.1 from HEASARC, with instrumental parameters
from {\sc caldb} updated 2008 September 5. Updates to the calibration
parameters released during the preparation of this paper do not affect
the results beyond the statistical uncertainties.

We performed the standard screening of events files to remove time
intervals during satellite manoeuvres, telemetry saturation, and
passage through the South Atlantic Anomaly, as well as elevation
angles less than $5\degr$ and $20\degr$ from the nighttime and daytime
Earth. The light curve for the remaining time intervals is stable,
showing no sign of flaring or excess particle background. We exclude
the corners of the detectors where $^{55}$Fe calibration sources lie,
and obvious point sources seen in \emph{Suzaku} or \emph{XMM} images were
excised with circular regions of radius $2.5'$ to remove more than 99
per cent of the flux spread by the PSF. Detector response matrices and effective area
functions are constructed with xisrmfgen and xissimarfgen
\citep{ishisaki07}, respectively. The latter tool generates the
spectral response for a given sky brightness, accounting for
the known issue of contamination on the optical
blocking filter. Uncertainties in this contamination should not
affect our results significantly, as we do not consider energies below
$1\kev$. Vignetting effects will be different for cluster and
background emission components, so we use a uniform surface brightness out to a radius of $20'$
when modeling the background emission and input a $\beta$-model
surface brightness profile using the parameters of \citet{chen03} for
the auxiliary response file applied to the cluster emission model. We
find that modifying the parameters of the surface brightness profile
used to determine the ``effective area'' with xissimarfgen does not
significantly impact the output. Thus, we do not expect an unresolved
central cusp or uncertainties from the outward extrapolation of the
surface brightness profile to influence our results.

\begin{table}
  \caption{Observational parameters of the five pointings}
  \label{obsdetails}
  \begin{center}
    \leavevmode
    \begin{tabular}{lllll} \hline \hline
    Obs. ID & Position & Exposure (ks) & RA & Dec (J2000) \\ \hline
    802062010 & Center & 32.0 & 116.8852 & -19.2901 \\
    802062020 & NW & 32.2 & 116.6543 & -19.2063 \\ 
    802062030 & NE & 30.8 & 116.9737 & -19.0727 \\
    802062040 & SE & 32.9 & 117.1155 & -19.3739 \\ 
    802062050 & SW & 33.4 & 116.7966 & -19.5079 \\ \hline
    \end{tabular}
  \end{center}
\end{table}

\subsection{Background Subtraction and Modelling}

Accounting for background emission is critical when observing regions
of low surface brightness, and it consists of multiple components. We
subtract the non X-ray background (NXB) of charged particles and gamma
rays with xisnxbgen \citep{tawa08}, which models the trend of these
events from night Earth data, weighted by the magnetic cutoff
rigidity. The X-ray background consists of solar wind charge exchange,
thermal emission mainly from the hot local bubble and the Galactic halo, and
the cosmic background (CXB) due to unresolved point sources. Charge exchange
contaminates the low energy spectrum with emission lines that can vary
on a time-scale of 10 minutes \citep{fujimoto07}, while the diffuse
thermal emission and CXB are expected to be more stable
and easily modeled. At the low Galactic latitude of this source
$(b=3\degr)$, the soft thermal background could vary significantly with
position. We opt to exclude the high energy end of the
spectrum where the NXB component dominates with several emission
lines, and to also ignore the low energy end where charge exchange and
local thermal emission muddle the data, leaving the $1-5\keV$ band for
consideration. Since the central regions have temperatures higher than
the upper limit of the energy band used, we test that they are
consistent with temperatures measured in a $1-8\kev$ band. Systematic
uncertainties due to particle background subtraction at high energies
and Galactic and other low energy background components are discussed
in Section~\ref{uncertainties}. We opt for the narrower energy range
to optimize the signal to noise for whole cluster.

Since cluster emission fills the observed field, we cannot use
detector regions from these pointings to subtract the remaining
background. Instead, we analyze \emph{Suzaku} data from the Lockman Hole
(observation ID 102018010) taken only 8 days prior to the start of our
observations. Despite the large difference in absorbing column density
measured from \hone{} maps \citep[$\nh=5.7 \times 10^{19}\cmsq$ for
the Lockman Hole versus $4.2 \times 10^{21}\cmsq$ for PKS
0745-191;][]{kalberla05}, \emph{ROSAT} observations \citep{snowden97}
detect similar background levels from $1-2\kev$ at the two
positions. We find that the Lockman Hole data from $0.5-8\kev$ is
well-modeled by a power law of photon index 1.4, absorbed by the
column density measured by the \hone{} maps, added to an unabsorbed
{\sc mekal} thermal plasma model \citep{mewe85,mewe86,liedahl95} at
$0.1\kev$ with solar metallicity using the relative abundances of
\citet{anders89}, allowing only the normalizations to vary. The best
fitting normalizations in the $1-5\kev$ range are
$7.26\times10^{-4}\pownorm$ at $1\kev$ for the CXB and $0.78\mknorm$
for the soft thermal component.\footnote{We follow \citet{sato07}
  in dividing the surface brightness normalization of the thermal
  component by the solid angle used in the ancillary response file
  generation, $\Omega=20'$, so that
  $Norm=10^{-20}\int{n_en_HdV}/\{4\pi(1+z)^2D_A^2\}/\Omega$ in
  $\mknorm$, where $D_A$ is the angular diameter distance to the cluster.}
We do not
find an improvement in the fit with another low-temperature component, as
is often used \citep[\emph{e.g.,} ][]{vikhlinin05}. We note that in
the $1-5\kev$ band used here, the fit is not even particularly
sensitive to the normalization of the remaining soft thermal component, but
we include it as an empirical description of the background model to
be applied to the PKS 0745-191 data. We attempted to fit the outermost
regions of the cluster observations with a similar model and an added
low-temperature ($\sim0.25\kev$) Galactic component. The
normalizations required are much higher than for the Lockman Hole and
would be inconsistent with the \emph{ROSAT} observations for this
source. The fit to the outer regions of PKS 0745-191 is improved with a higher
temperature component ($\gtrsim1.5\kev$), indicative of cluster
emission covering the entire field.

\section{Spectral Analysis}
\label{analysis}

Spectra were extracted in annular regions of
radii $0'-2.5', 2.5'-6', 6'-9.5', 9.5'-13.5', 13.5'-18.5'$, and
$>18.5'$, with the outermost region reaching to nearly $24'$. For each
annulus we define an effective radius that is approximately an
emission-weighted mean \citep{mclaughlin99}, $r=[0.5(r_{\rm
      in}^{3/2}+r_{\rm out}^{3/2})]^{2/3}$. These regions
are centered at right ascension and declination 07:47:31.325,
-19:17:39.95 (J2000), which is coincident with the cD galaxy and both the
peak and centroid of X-ray emission. Due to the low count rates in
some channels, we use the \citet{cash79} C statistic.

We used {\sc xspec} v12.5.0 to simultaneously fit models of the
spectra, with data from each pointing grouped by annulus unless
specified otherwise. We first
subtract the NXB spectra from identical regions on the detector, scaled by the
relative exposure time of the night Earth observations. Next, we
model the local thermal emission and CXB components, fixed
to the best-fitting values from the Lockman Hole observations. Finally, we
model the remaining emission, that from the cluster, with another
{\sc mekal} component at the cluster's redshift, $z=0.1028$. We fix
the column density absorbing the cluster and CXB components to the
best-fitting value at the center, $\nh=3.6 \times 10^{21}\cmsq$, which
is lower than the \hone{} measurement but consistent with the value
from \emph{XMM}'s EPN \citep{chen03}. We allow the metallicity to vary
in the center but fix it to $0.3\zsun$ at larger radii. Of the
remaining parameters, only the temperature and normalization of the
thermal ICM component are allowed to vary. 

\begin{figure}
  \begin{center}
    \leavevmode
      \epsfig{figure=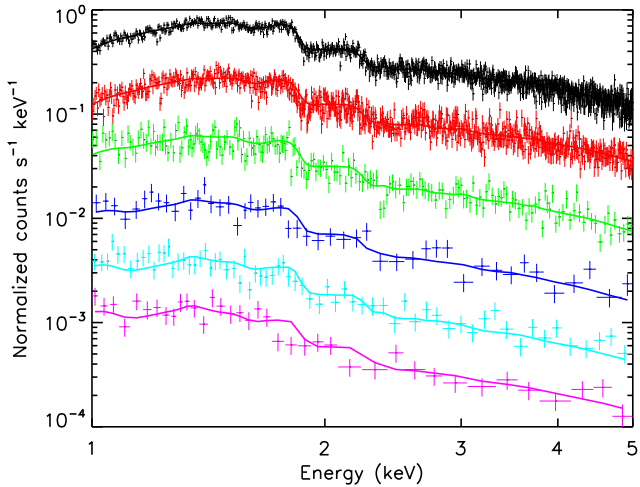,
        width=\linewidth}
      \caption{Data and spectral fits including PSF-correction for
        each annular region, illustrating the softening of the spectra
        with increasing radius. The regions plotted are
        $0'-2.5'$ (black), $2.5'-6'$ (red), $6'-9.5'$ (green),
        $9.5'-13.5'$ (blue), $13.5'-18.5'$ (cyan), and $>18.5'$
        (magenta). Data have been rebinned to a minimum
        significance of $5\sigma$ for each point, and each spectrum has
        been arbitrarily renormalized for visual clarity.}
      \label{allspec1}
  \end{center}
\end{figure}

\begin{figure}
  \begin{center}
    \leavevmode
      \epsfig{figure=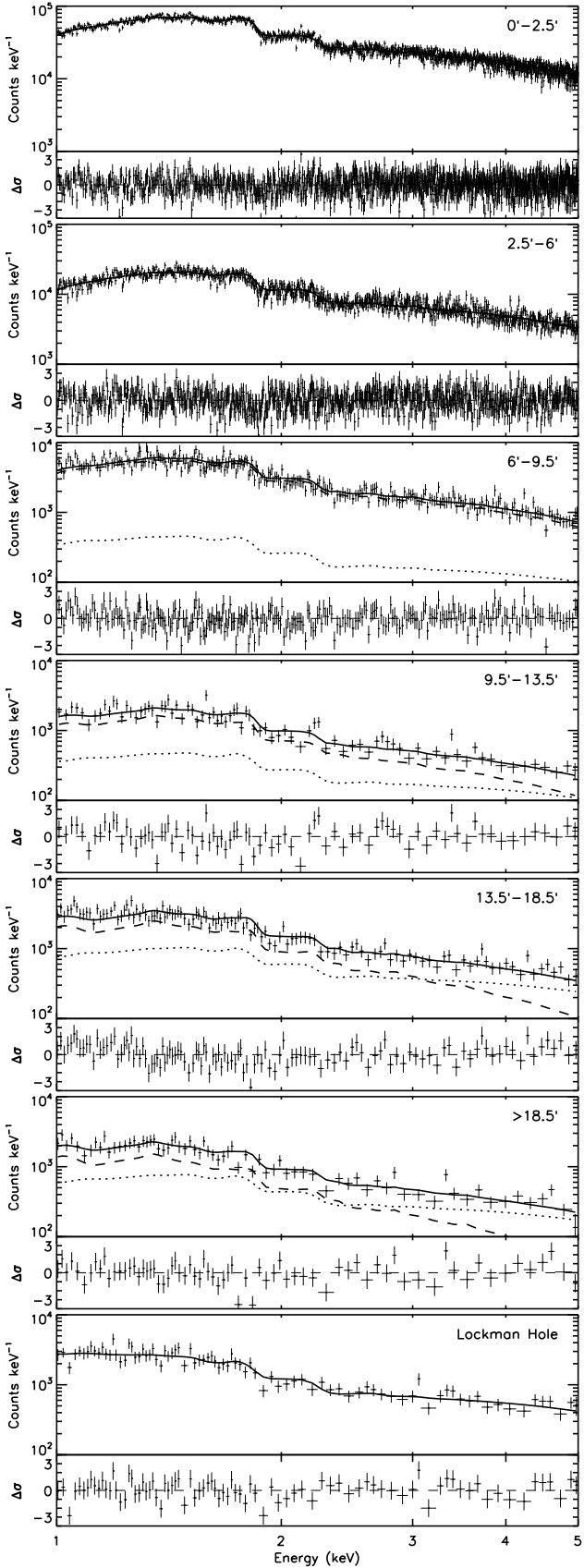,
        width=\linewidth}
      \caption{Same as Fig.~\ref{allspec1}, but with residuals to show
        the quality of the spectral fits, and with the Lockman Hole
        background spectrum included for comparison. Count levels have
        not been renormalized and we note that the regions have
        different areas and exposure times. Dotted and dashed curves
        represent the background (CXB+thermal) and cluster emission
        models, respectively.}
      \label{allspec2}
  \end{center}
\end{figure}

The broad point-spread function (PSF) of \emph{Suzaku}, with a
half-power diameter of $\sim2'$, will cause some
emission from each annulus on the sky to be distributed into others on
the detector. We correct for this effect following the method of
\citet{sato07}, who importantly show that \emph{Suzaku}'s PSF is nearly
energy-independent. We use a circularly symmetric $\beta$-model with the
best-fitting parameters of \citet{chen03}, consistent with the surface
brightness profile observed here, and a monochromatic energy of
$1.5\kev$ to generate a photon list that is propagated with a
ray-tracing simulator (xissim) through the detector optics. For each
annulus on the sky, we use the ratio of flux landing in the
corresponding detector annulus to the flux in each other annulus to
calculate cross-annuli normalization factors. In the spectral model,
we add a thermal component from each annulus to every other annulus
with the temperature tied to the original region and the normalization
scaled by these factors. The relative contributions to each annulus
from other annuli, as derived from these PSF correction factors and
the best fitting normalizations, are given in
Table~\ref{psfnorms}. The model was fitted simultaneously over all 
annuli and resulted in an excellent fit; the C statistic, though not a
direct measure of goodness of fit, was 65413 using 98640 bins and
98627 degrees of freedom. The spectra and models are shown in
Figs.~\ref{allspec1} and \ref{allspec2}. We note that the individual
spectra are fitted simultaneously, and are only combined by annular
regions in these figures for visual clarity.

\begin{table*}
   \begin{center}
    \leavevmode
    \begin{tabular}{lrrrrrr} \hline \hline
     & 0'-2.5' & 2.5'-6' & 6'-9.5' & 9.5'-13.5' & 13.5'-18.5' & 18.5'-24' \\ \hline
    0'-2.5' & 89.9 & 19.3 & 2.6 & 0.4 & $<0.05$ & 0.1 \\
    2.5'-6' & 9.7 & 66.8 & 5.7 & 0.6 & 0.1 & 0.1 \\ 
    6'-9.5' & 0.3 & 13.1 & 84.4 & 7.8 & 0.6 & 0.1 \\
    9.5'-13.5' & 0.1 & 0.5 & 6.5 & 78.5 & 7.3 & 0.8 \\
    13.5'-18.5' & $<0.05$ & 0.1 & 0.5 & 11.9 & 84.1 & 10.9\\
    18.5'-24' & $<0.05$ & 0.1 & 0.3 & 0.7 & 7.8 & 88.1 \\ \hline
    \end{tabular}
  \caption{Relative flux contributions due to PSF spreading. Values
    correspond to the fraction of flux measured in the detector region
  (columns) originating from the cluster region (rows), \emph{i.e.,}
  of the flux measured in the outermost annulus, we expect $88.1\%$ to
  come from the corresponding region in the sky, and only $0.1\%$ to
  have originated in the central projected region.}
  \label{psfnorms}
  \end{center}
\end{table*}

\section{Results}
\label{results}

\subsection{Temperature, density, and entropy profiles}

\begin{figure}
  \begin{center}
    \leavevmode
      \epsfig{figure=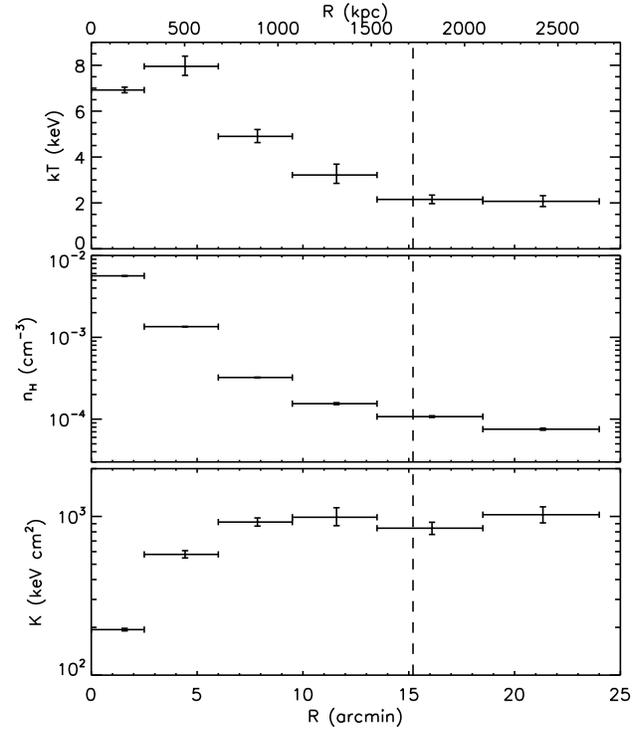,
        width=\linewidth}
      \caption{Projected temperature, density, and entropy
        profiles. The vertical dashed line shows our estimate of
        $r_{200}$, derived in Section~\ref{massprofilesection}.}
      \label{profiles}
  \end{center}
\end{figure}

Temperature and density profiles are the primary data products of
X-ray observations of clusters, from which other properties including
entropy, mass, and gas fraction profiles can be derived. We obtain the
density from the normalization of the best-fitting thermal plasma
model which is proportional to the emission measure, 
$\int{n_{\rm e}n_{\rm H}\,dV}$, where $n_{\rm e}\approx 1.2n_{\rm H}$
for typical abundances. We use the simpifying assumption of
constant density within each annulus, and estimate the volume
as that of the intersection between cylindical and spherical shells
with shared inner and outer radii, $V=(4\pi/3)(r_{\rm out}^2-r_{\rm
  in}^2)^{3/2}$, scaled by the fraction of the projected annulus
observed. With the density and temperature determined, we can 
calculate the entropy profile, $K=kT/n_{\rm e}^{2/3}$. Each of these profiles
is shown in Fig.~\ref{profiles}.

The cool core, though blurred by the wide PSF, is seen as a decrease
in temperature near the center. The more novel result is the
clearly-observed decline in temperature in the outskirts of the
cluster. Excluding the core, this decline can be fit with a power law
$T(r)=T_0r^{\alpha_T}$ with index $\alpha_T=-0.94\pm0.06$. The gas
density profile can similarly be fit with a power law, excluding the
central region again for uniformity, with index
$\alpha_n=-2.23\pm0.01$. If we assume a polytropic distribution for an
ideal gas, $P\propto n_{\rm H}kT \propto n_{\rm H}^{\Gamma}$, then we find
$\Gamma=1+(\alpha_T/\alpha_n)=1.42\pm0.03$. This value is higher than
that found for cooling flow clusters determined by \emph{e.g.,} 
\citet{degrandi02}, who measured $\Gamma\simeq1.2$ in the radial range
$0.2r_{180} < r \lesssim0.6r_{180}$.

Gas with high entropy rises, so the radial entropy profile is expected
to increase. Puzzlingly, the entropy profile we observe rises from the center
and then appears to level off beyond $\sim10'$, putting it on the
border of convective instability. We will discuss possible
explanations for this behavior in later sections.

\subsection{Mass profiles}
\label{massprofilesection}
To derive a mass profile, we assume that the gas properties are
spherically symmetric and that the cluster is in hydrostatic
equilibrium. Balancing the thermal pressure gradient against the
gravitational potential, we obtain an expression for the total
gravitating mass in terms of temperature and density:
\begin{equation}
\label{hydroeq}
M(<r)=\frac{-kr^2}{G\mu m_{\rm H} n_{\rm H}}\left(T\frac{dn_{\rm
    H}}{dr} + n_{\rm H}\frac{dT}{dr}\right).
\end{equation}
We assume that the total density distribution can be described by an
NFW profile, 
\begin{equation}
\rho(r)=\frac{\rho_0}{r/r_{\rm s}\left(1+(r/r_{\rm s})\right)^2},
\end{equation}
where $\rho_0$ is a density normalization and $r_{\rm s}$ is a scale
radius related to the cluster's concentration, $c_{200} \equiv
r_{200}/r_{\rm s}$. 

Following an approach similar to that of \citet{schmidt07}, we
use an NFW model and the observed gas density distribution to predict
a temperature in each annulus, beginning with the outermost
bin. Density gradients are taken linearly between the radial bins,
which reduces the correlation in uncertainties between non-adjacent
data points that are created with other interpolations such as cubic
splines, as some authors use \citep[see ][ for a
discussion]{voigt06}. This approach 
could be improved with more radial bins, but the wide PSF of
\emph{Suzaku} and the low count rate in the cluster outskirts limit us
from using thinner annular regions. To ensure that the outermost
temperature value is robustly estimated, we take the median outer
value of the profiles that best fit 100 Monte Carlo realizations of
the observed temperature profile, normally distributed within its
error bars. The model temperature profiles are similar if we begin
at the innermost region, so the result is not particularly sensitive
to the boundary values. Iterating over a range of NFW parameters, we
select the mass model that produces the temperatures that best fit the
observed profile.

In order to resolve the region within the NFW scale radius, we
supplement the \emph{Suzaku} data with an archival $20\ks$
\emph{Chandra} observation (ID 2427) taken with the ACIS-S detector in VFAINT
mode. These data were reduced following the description of \citet{odea08},
with spectra extracted in circular annuli and deprojected using the method of
\citet{sanders07}. After using the outermost \emph{Chandra} annulus to
subtract contaminating flux from inner regions it is omitted from
further analysis since it may contain external cluster emission,
leaving six regions to replace the innermost \emph{Suzaku} bin, where
the temperature and density is in reasonable agreement with the
\emph{Chandra} data. We use \emph{Chandra} data for its sharper PSF,
but the \emph{XMM} temperature profile in \citet{snowden08} is also
consistent at the smaller radii of that observation.

\begin{figure}
  \begin{center}
    \leavevmode
      \epsfig{figure=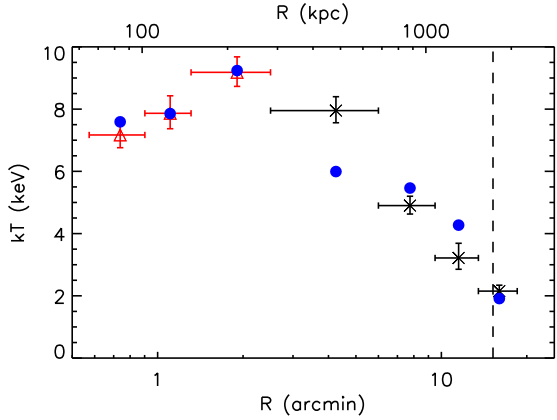,
        width=\linewidth}
      \caption{\emph{Chandra} (red triangles) and \emph{Suzaku} (black
        crosses) temperature values used in mass fitting
        analysis, along with matching temperatures for best fitting
        NFW profile (blue circles).} 
      \label{nfwtemp}
  \end{center}
\end{figure}

The model temperature profile, shown in Fig.~\ref{nfwtemp}, does not fit the data very well over all
annuli, and the best fit to the observed temperatures
($\chi^2\approx39$ for 7 degrees of freedom) is obtained after
ignoring the outermost \emph{Suzaku} bin which is
beyond the virial radius where the assumption of hydrostatic equilibirium is
expected to fail, and the \emph{Chandra} regions in the cool
core within $\sim70\kpc$ of the center where the gas is likely to be
multiply-phased. The main deviation in our data
from this acceptable temperature fit arises from the $2.5'-6'$
\emph{Suzaku} annulus. Possible explanations include a PSF-correction
that could be too large, a cross-calibration issue between
\emph{Chandra} and \emph{Suzaku}, or even non-gravitational heating in
that region.

\begin{figure}
  \begin{center}
    \leavevmode
      \epsfig{figure=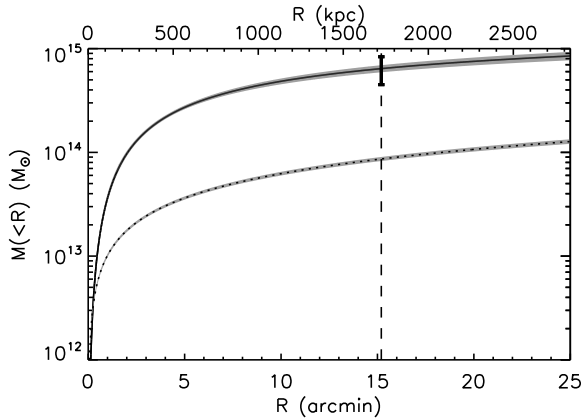,
        width=\linewidth}
      \caption{Cumulative profiles of gas mass (dotted lower curve) and total
        gravitational mass (solid upper curve) with $1\sigma$ statistical
        error ranges in gray. The single error bar shows the $30\%$ systematic
        uncertainty on $M_{200}$ and the vertical dashed line
        shows our estimate of $r_{200}$, derived in
        Section~\ref{massprofilesection}.} 
      \label{massprofile}
  \end{center}
\end{figure}

The resulting mass profile has best-fitting NFW parameters of
$c_{200}=7.5^{+1.1}_{-0.9}$ and $r_{\rm s}=230^{+40}_{-40}\kpc$,
producing a virial radius $r_{200}=c_{200}r_{\rm s}
=1.72\pm0.06\mpc$, or $15.2'$ and mass $M_{200} =
200\rho_c(z)(4\pi/3)(r_{200})^3
=6.4\pm0.6\times10^{14}\msun$ . We estimate the total uncertainty on
the virial mass, including systematics, of $30\%$, which is described
further in the next section. We compare these values with
previous determinations in Table~\ref{massparams} and plot the mass
profile in Fig.~\ref{massprofile}, along with the cumulative gas
mass profile obtained from the power-law fit to the \emph{Suzaku} gas
density profile.

\begin{table}
   \begin{center}
    \leavevmode
    \begin{tabular}{lll} \hline \hline
    $c_{200}$ & $M_{200} (10^{14}\msun)$ & Reference \\ \hline
    $3.83^{+0.52}_{-0.27}$ & $18.6^{+3.5}_{-4.0}$ & \citet{allen03} \\
    $5.12^{+0.40}_{-0.40}$ & $10.0^{+1.2}_{-1.2}$ & \citet{pointecouteau05} \\ 
    $5.46^{+3.22}_{-2.88}$ & $9.7^{+52.2}_{-8.5}$ & \citet{voigt06} \\
    $5.86^{+1.56}_{-1.07}$ & $11.82^{+4.70}_{-3.55}$ & \citet{schmidt07} \\
    $7.5^{+1.1}_{-0.9}$ & $6.4^{+0.6}_{-0.6}$ & This work \\ \hline
    \end{tabular}
  \caption{Concentration and virial mass determinations for PKS
    0745-191, compiled by \citet{comerford07} with values for this
    work added.}
  \label{massparams}
 \end{center}
\end{table}

The NFW concentration is higher than previous estimates, and the
virial mass we determine is lower, due in part to the
smaller virial radius found in this work. \citet{reiprich02}, who used
an isothermal fit to \emph{ROSAT} and \emph{ASCA} data and found a
value of $1.95\pm0.03\mpc$ updated to the cosmological parameters we
adopt. \citet{schmidt07} found an even larger value for the virial
radius, fitting an NFW profile to \emph{Chandra} data in the central
region to obtain $r_{200}\approx2.2\mpc$. One issue with past NFW fits used
to calculate the virial radius is that the data often do not extend
even to the NFW scale radius, let alone the virial radius. 

We can compare our mass profile with the projected mass determined
from a strong lensing arc of radius $34.5\kpc$ seen with \emph{Hubble} imagery by
\citet{allen96}. Updating their mass estimate for the circular lens
model to account for the current cosmological parameters, we expect a
mass of $6.4\times10^{12}\msun$ projected within a cylinder along the
line of sight of the center of the cluster with a radius of the
lensing arc, which is consistent with the statistical and 
systematic uncertainties of our best-fitting NFW model when
integrating the mass in the cylinder out to $r_{200}$. We note that this includes
an extrapolation of our profile into the core region where we have
excluded data from our fit because of nonthermal processes. While it
is reassuring to see this reasonable agreement between separate mass
estimates, we caution that comparisons of this type of wide-field
X-ray analysis with strong lensing data only involve a small fraction
of the total cluster mass. Weak lensing analyses, which can provide
mass projections over a much broader area, will provide important
comparisons for virial mass estimates.
 
\begin{figure}
  \begin{center}
    \leavevmode
      \epsfig{figure=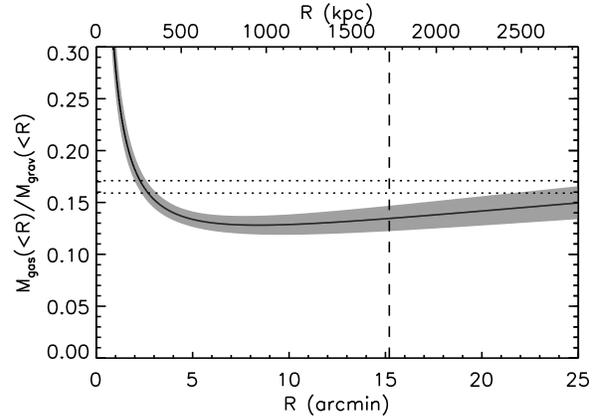,
        width=\linewidth}
      \caption{Gas fraction profile for PKS 0745-191, the ratio of the
      dotted to solid curves in Fig.~\ref{massprofile}, with $1\sigma$
      statistical error range in gray. Dotted lines
    show the $1\sigma$ range around the value of $\Omega_{\rm
      b}/\Omega_{\rm m}$ from \citet{komatsu08}. The vertical dashed
    line shows our estimate of $r_{200}$, derived in
    Section~\ref{massprofilesection}.} 
      \label{gasfraction}
  \end{center}
\end{figure}

In Fig.~\ref{gasfraction}, we plot the gas fraction profile, defined
as the fraction of mass within a given radius composed of X-ray
emitting gas. The gas fraction appears to rise slightly at the outer radii, consistent
with the profile observed for PKS 0745-191 in \emph{ROSAT} data by
\citet{allen96}. The values of the $f_{\rm gas}$ profile in the
outskirts of this cluster are similar to recent measurements of the mean
cosmic baryon fraction \citep{komatsu08}.

\subsection{Uncertainties}
\label{uncertainties}

\begin{table*}
   \begin{center}
    \leavevmode
    \begin{tabular}{lcccccccc} \hline \hline
    Radius & kT & Stat. & No PSF & NXB & CXB & Gal. & $\nh$ & Z \\ 
    (') & ($\kev$)& & & $\pm3\%$ & $\pm10\%$ & $\pm50\%$ & $\pm10\%$ & $0.2-0.4\zsun$ \\ \hline
    0--2.5     & 6.92 & $^{+0.12}_{-0.12}$ & $+0.08$ & $<0.005$ & $<0.005$ & $<0.005$ & $0.68$ & $<0.005$ \\ 
    2.5--6.0   & 7.85 & $^{+0.45}_{-0.39}$ & $-0.82$ & $<0.005$ & $<0.005$ & $0.01$ & $0.90$ & $0.02$ \\ 
    6.0--9.5   & 4.90 & $^{+0.30}_{-0.27}$ & $-0.01$ & $0.02$ & $0.04$ & $0.03$ & $0.41$ & $0.05$ \\ 
    9.5--13.5  & 3.22 & $^{+0.47}_{-0.36}$ & $-0.05$ & $0.14$ & $0.14$ & $0.06$ & $0.28$ & $0.07$ \\ 
    13.5--18.5 & 2.15 & $^{+0.19}_{-0.18}$ & $+0.09$ & $0.06$ & $0.17$ & $0.06$ & $0.15$ & $0.11$ \\ 
    $>18.5$   & 2.07 & $^{+0.25}_{-0.23}$ & $+0.02$ & $0.10$ & $0.21$ & $0.06$ & $0.13$ & $0.12$ \\ \hline
    \end{tabular}
    \caption{Estimated uncertainties in the radial temperature profile
      due to removing PSF-corrections or changing normalizations of
      background components and spectral model parameters within the
      ranges given. Each parameter is varied independently and we
      present only the maximum of the  upward and downward temperature
      shifts (in $\kev$) for clarity. Best-fitting PSF-corrected
      temperatures and their statistical uncertainties are included
      for comparison.}
  \label{errors}
 \end{center}
\end{table*}

A number of systematic uncertainties could add contributions to our
error budget beyond the statistical ranges plotted. We summarize our
estimates of several of these systematic uncertainties to the
temperature profile, including the
effects of PSF-correction and renormalization of background
components, in Table~\ref{errors}. The changes due to the
PSF-correction are mostly small, with the main effect being a $\sim12$
per cent increase in temperature in the second innermost annulus after
accounting for emission from the cool gas in the core that has
scattered outward. There is a significant uncertainty in the
temperature of the central region because \emph{Suzaku}'s PSF blurs
the steep gradient of the cool core. As \citet{reiprich08} points out,
the approach to PSF-corrections described in the previous section does
not fully account for this blurring within the central region, however
our method provides a reasonable approximation to the
corrections needed. 

To ensure that the background models used do not significantly impact
the results, we vary the normalizations of these components and
measure the deviations produced in temperature. The
uncertainty in the NXB is $\sim3$ per cent \citep{tawa08}, and from an
analysis of \emph{ROSAT} data by \citet*{carrera97}, we estimate that
the cosmic variance due to point source clustering in the CXB is $\sim10$
per cent over this field of view. We also test the effect of
metallicity from $0.2-0.4\zsun$ in the outer annuli and a range of
$\pm10$ percent in the column density. Each of these variations results in
a change in temperature of $\lesssim10$ per cent for all annuli, and
the individual systematic uncertainties are smaller than the
Poisson uncertainties for all but the column density's effect on the
central regions where the count rate is high. Finally, the
normalization of the local thermal component, which could vary the
most across the cluster field of view because of the low Galactic
latitude, actually has a very small effect on temperature fits above $1\kev$.

Without spectral deprojection, the data from each annulus will be
contaminated by emission from regions at larger radii. But because the
density profile falls so steeply and the thermal bremsstrahlung emission
scales as the square of the density, the only noticeable effect is in
the cluster core \citep[\emph{e.g.,} Fig. 3 of ][]{sanders07}. This region is
well within our innermost radial bin, so we do not expect significant
errors due to the lack of deprojection of the \emph{Suzaku} data.

\begin{figure}
  \begin{center}
    \leavevmode
      \epsfig{figure=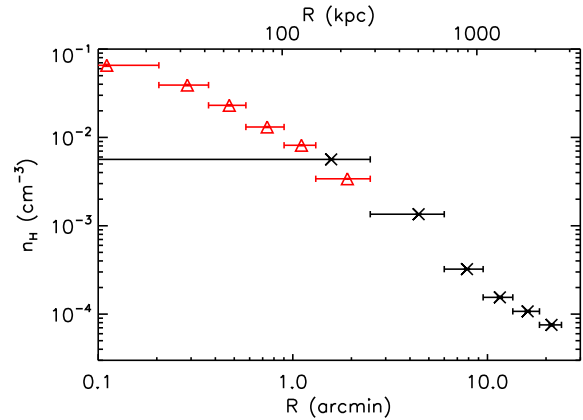,
        width=\linewidth}
      \caption{Comparison between \emph{Chandra} (red triangles) and
        \emph{Suzaku} (black crosses) density profiles. Error bars are
      smaller than the plot symbols.}
      \label{density_chandra}
  \end{center}
\end{figure}

As an additional check on the reliability of our temperature and
density profiles, we compare the values from \emph{Suzaku} with
the \emph{Chandra} observations used in the mass estimate. The high
spatial resolution of \emph{Chandra} allows for our central region to
be divided into several annuli, with the average temperature in
agreement with the \emph{Suzaku} value. As shown in
Fig.~\ref{density_chandra}, the density profiles are also reasonably
consistent between the two observatories, with a slightly steepening
slope at the larger radii seen by \emph{Suzaku}. 

\begin{figure}
  \begin{center}
    \leavevmode
      \epsfig{figure=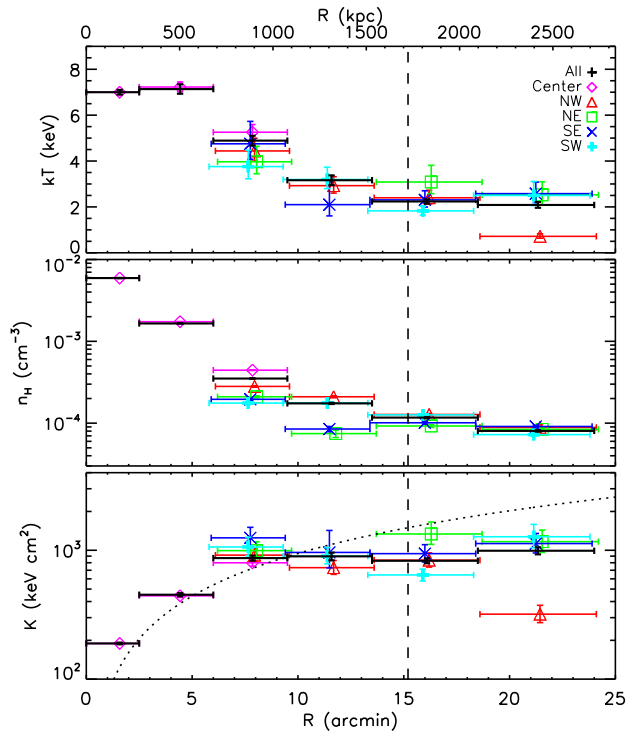,
        width=\linewidth}
      \caption{Comparison of temperature, density, and entropy
        profiles in each direction. The temperature in the NE
        direction is not well-constrained from $9.5'-13.5'$ due to a
        large amount of area excluded because of point sources, so the
      temperature and entropy points are not plotted in this
      region. Slight offsets are added to bin radii for clarity. The
      dotted curve in the bottom panel shows $K \propto r^{1.1}$. The
      vertical dashed line shows our estimate of $r_{200}$, derived in
      Section~\ref{massprofilesection}.} 
      \label{profiles_4dir}
  \end{center}
\end{figure}

We can check the assumption of spherical symmetry used when deriving
the mass profile, seeing if the emission is at least circularly
symmetric by considering the projected profiles in each of the
pointings separately. Fig.~\ref{profiles_4dir} shows the
temperature, density, and entropy profiles for each pointing. Here we
do not include PSF-corrections in order to keep each set of observations
independent. The uncertainties are larger when splitting up the data,
but the results from the separate pointings broadly overlap, and no
single direction is consistently above or below the average
temperature or density for every annulus. Scatter that is larger than
the statistical uncertainties could be attributed to remaining point
sources or minor asymmetry in the cluster. 

The entropy seen in the outermost radial bin of the NW pointing is
significantly lower than nearby regions, mainly due to the temperature
drop seen there. However, scatter between pointings or
individual outliers are unable to explain the observed flattening of
the entropy profile in the summed annuli. We plot a power law rising
as $r^{1.1}$ to compare with the entropy profile predicted by
analytical models of accretion shock heating and seen in some
observations as well as numerical simulations \citep{tozzi01,
  ponman03, voit05entropy}. A small number of objects in the sample of
\citet{ponman03} have entropy curves that do not rise with
radius. The observed and expected profiles agree at small radii,
rising out of the core, but there is a clear entropy deficit in the
outskirts of the cluster, perhaps indicative of infalling gas which is
not dynamically stable. Our background model would have to be off by
an order of magnitude, or the column density changed by a factor of 2,
to increase the entropy in the outskirts to the predicted values.

\begin{figure}
  \begin{center}
    \leavevmode
      \epsfig{figure=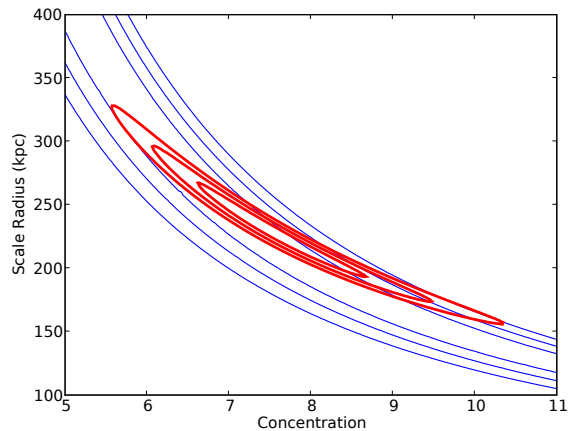,
        width=\linewidth}
      \caption{Likelihood contours for the concentration and scale
        radius parameters of the NFW fits, drawn at
        $\Delta\chi^2=2.30$, 6.17, and 11.80, corresponding to the
        $1\sigma$, $2\sigma$, and $3\sigma$ confidence intervals,
        respectively. Thick red contours show the constraints from the
        combination of \emph{Chandra} and \emph{Suzaku} data, excluding the
        cluster core and the outermost annulus which is beyond the virial
        radius. Thin blue contours show the constraints
        without the \emph{Suzaku} data for comparison.} 
      \label{nfwcontour}
  \end{center}
\end{figure}

For the estimates of the virial radius and mass, the method of fitting
an NFW model to our temperature profile is an additional source of
uncertainty. We have excluded the inner and outer regions where the
assumption of hydrostatic equilibrium is likely to break down, but several
variations of the temperature profile with these points included or
excluded produce values for $r_{200}$ consistent to $\sim10$ per
cent. Similar values are also obtained when using only the
\emph{Suzaku} data, which has a large uncertainty in the central
temperature due to PSF-blurring. The wide radial span of this data set
provides significantly tighter constraints on the NFW parameters than
could be obtained with the \emph{Chandra} data alone, as shown in
Fig.~\ref{nfwcontour}. Data from the cluster core are still excluded in the
contours plotted, though they might be used in a typical analysis. In
either case with only \emph{Chandra} data, the NFW scale radius is not
well-constrained since the data do not extend out that far, and the
resulting estimate for $r_{200}$, and thus $M_{200}$, would be
significantly larger than the values we obtain. We can also include the
systematic uncertainties in the temperature profile within this
process of fitting over NFW parameter space. Even with differences of
$0.5-1\kev$ between the FI and BI sensors which have similarly shaped
but offset temperature profiles, changes in the best fitting value of
$R_{200}$ are of order $10\%$, giving a systematic uncertainty of
$30\%$ for the determination of $M_{200}$. 

We can also estimate the cluster mass independently of any model
parametrization by plugging our density and temperature profiles
directly into Equation~\ref{hydroeq} and calculating gradients from
the finite difference quotients of adjacent points. At large radii,
the cumulative mass estimate using this method actually decreases,
indicating that the hydrostatic assumption fails, and the
uncertainties are large because spatial gradients are taken between points
separated by the wide thickness of annular bins. However, at the virial
radius near $15'$, the mass estimate agrees with that derived from the
NFW model to within $\sim30$ per cent.

\section{Discussion}
\label{discussion}

These are the first well-constrained measurements of the gas temperature
profile beyond $r_{200}$ for a rich and relaxed cluster, which we can now
compare with model predictions and simulations. We see
that there is a clear decline in temperature between $0.3-1.5
r_{200}$, consistent with recent observational trends at growing radii
\citep{vikhlinin05,pratt07,reiprich08} and suggesting that conduction
does not play a significant role on typical cluster
time-scales. \citet{roncarelli06} measured the magnitude of the
temperature drop in hydrodynamical simulations, finding a 40
per cent decline from $0.3-1 r_{\rm 200}$, the same size found in the
analytical treatment of \citet*{ostriker05}. The drop observed in PKS
0745-191 over this radial range is closer to 70 per cent, but the
general shapes of the profiles are similar.

The observed entropy profile deviates from expectations near the
virial radius. \citet{tozzi01} describe the effects on entropy of the
accretion phases, including adiabatic infall, shock heating, and
adiabatic compression followed by cooling within the
halo. Interestingly, we see evidence for an accretion shock beyond the
virial radius in one direction (NW pointing in
Fig.~\ref{profiles_4dir}), suggestive of cool material falling
inward along a filament. The change in entropy in the outer
annulus of this region is not seen in the other directions, though the
values there are still lower than would be expected from the typical
increase of $K\propto r^{1.1}$.

The sharper decline in temperature than predicted and the flattening
of the entropy profile in the outskirts suggest a need for nonthermal
pressure support in order to maintain dynamic stability. The numerical
simulations of \citet*{eke98} show a significant rise in the ratio of
bulk kinetic energy to thermal energy beginning within the virial radius. Merger
activity increases the turbulent pressure, and even though PKS
0745-191 appears relaxed morphologically, it should still be accreting
matter in the outskirts. We note that the infall time-scale, $t_{\rm f}\sim
(r^3/GM)^{1/2}$, is of order $1\gyr$ at the virial radius, and the
time-scale set by the speed of sound, $t_{\rm s}\sim (r^2m_{\rm H}/
\Gamma kT)^{1/2}$, is several times larger.

\citet{neumann05}, studying the summed profiles of 14 nearby Abell
clusters beyond $r_{200}$, argues that the ICM at large radii may not
be in hydrostatic equilibrium, and that cool gas not seen in X-rays
could increasingly dominate the baryon content in the
outskirts. \citet{afshordi07} stacked WMAP observations of a
large sample of massive clusters and found a deficit in thermal energy
in the outskirts from the Sunyaev-Zel'dovich profile, also arguing for
cool phase of the ICM. It is not well understood how these different
phases would mix, and complicated gas physics would likely result. The
assumption of hydrostatic equilibrium fails beyond the virial radius,
as seen by the poorer temperature fits from mass modeling when the outermost
temperature value is included. It is not obvious to what extent the ICM
\emph{within} the virial radius is not in equilibrium, but the entropy
and temperature profiles suggest that nonthermal physics may be
important in the outskirts of this cluster. 

\begin{figure}
  \begin{center}
    \leavevmode
      \epsfig{figure=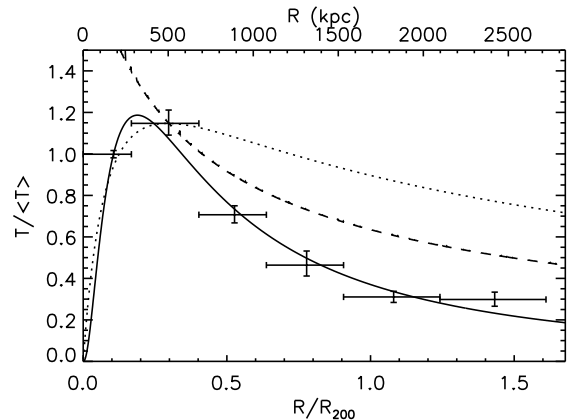,
        width=\linewidth}
      \caption{\emph{Suzaku} temperature profile, normalized by the
        emission-weighted mean and compared with model profiles. The
        solid curve shows the parametrization of
        Equation~\ref{newfit}, also rescaled by the
        emission-weighted mean temperature that we observe. The dotted
        curve shows the temperature for gas tracing an NFW
        distribution, while the dashed curve shows a polytropic
        profile in hydrostatic equilibrium with an NFW potential. Both
        models are normalized to match the maximum observed temperature.} 
      \label{tempmodels}
  \end{center}
\end{figure}

In Fig.~\ref{tempmodels}, we plot the observed temperature profile
against two models. The first is the expected temperature if the gas
content traced the NFW distribution of dark matter, with a 
Keplerian velocity profile, \emph{i.e.,} $T(r)\propto v(r)^2\propto
M(<r)/r$. The second and more realistic case is for the projected
X-ray temperature of gas arranged polytropically and in hydrostatic
equilibrium with an NFW gravitational potential, as derived by
\citet*{suto98}. The outer temperatures we have observed are lower than
those expected for reasonable values of the model
parameters\footnote{We set the parameter $B_{\rm p}$, defined by
  \citet{suto98}, equal to unity, and use the polytropic index derived
  from the best-fitting power laws to our data excluding the central
  region, $\Gamma=1.5$. It is possible to obtain temperature profiles
  that decline as steeply as the observed one by increasing the value
  of $B_{\rm p}$, though this would require truncating the gas
  distribution shortly beyond the regions we have measured. For the
  range $\Gamma=1.1-2$, the polytropic index used does not have a
  significant effect on the shape of the scaled
  profile.}. Additionally,  we can fit the observed temperature
profile using a simple parametrization 
\begin{equation}
\label{newfit}
T(r)=T_1\left(\frac{r}{r_s}\right)^\alpha\left(1+\frac{r}{r_s}\right)^\beta
\rm keV
\end{equation}
with best-fitting parameters $T_1=135$, $\alpha=2.4$, $\beta=-4.1$,
where $r_{\rm s}$ is the scale radius from the best-fitting NFW
profile. We note that the normalization is related to the value of the
temperature at this radius, $T_1=T(r_{\rm s})/2^\beta$.

A possible explanation for the deficit of thermal energy seen at large radii is
that infalling matter has retained some of its kinetic energy in bulk
motion. If there is increasing pressure support from bulk motion,
turbulence, or other nonthermal processes in the outer regions
of the cluster, then the gravitational potential and thus total mass would be
incorrectly estimated when assuming hydrostatic equilibrium. This
effect would depend on the changes in both the temperature and its
gradient, making estimation of the true mass difficult. 

Alternatively, the peak temperature, seen in the second-innermost
annulus, could be inflated due to nonthermal processes such as shock
heating during infall. AGN heating is thought to play an important
role in cool core clusters, and some excess energy could also heat the
gas beyond the core. More observations are needed of this cluster and
others to determine how well the self-similar scaling relations, which
depend on the dominance of gravitational process in creating
equilibrium, apply out to the virial radius.

The gas fraction we obtain is more consistent with previous
results. At $r_{200}$, $f_{\rm gas}=0.13$, lower than that found with
\emph{XMM} \citep{chen03}, but similar to the \emph{ROSAT} value of
\citet{allen96}. The trend of $f_{\rm gas}$ rising with radius is also
seen in other clusters and in simulations \citep[\emph{e.g.,}
][]{vikhlinin06,eke98}, since the gas density distribution is less
centrally concentrated than that for collisionless dark matter. The
outermost measured value of $f_{\rm gas}$ is consistent with the mean cosmic
baryon ratio derived from \emph{WMAP} data and other cosmological
observations, $\Omega_{\rm b}/\Omega_{\rm m}=0.165\pm0.006$
\citep{komatsu08}.

\section{Summary}
\label{summary}

We have presented the first measurements of the gas beyond the virial
radius from a rich and apparently relaxed galaxy cluster. The
temperature profile shows a significant decline at outer radii, while
the entropy profile levels off. The spatially resolved temperature and
density measurements out to and beyond the virial radius produce improved
constraints on the mass and gas fraction profiles, though the observed
temperature profile does show deviations from that expected from
cluster gas in hydrostatic equilibrium with an NFW mass profile.

Observations of the outskirts of clusters offer a direct probe of
their assembly history. The apparent shock front in one direction
makes it clear that this cluster is still accreting material, likely
along a filament. Evidence for nonthermal pressure support suggests
that bulk motions from merger activity could be making a significant
contribution to the gas energy in the outskirts of this cluster.

PKS 0745-191 is an ideal candidate for such observations, given its
distance and luminosity. The angular size fits reasonably within a few
pointings and the count rate is high enough for good spectral
statistics. The low and well-constrained background levels of
\emph{Suzaku} are crucial for low surface brightness
measurements like those presented here and by \citet{reiprich08}. We
expect that as the number of cluster observations in the \emph{Suzaku}
archive increases, these types of analyses will be able to probe the
self-similarity of the outskirts of clusters. More studies at the
virial radius, including X-ray observations, gravitational lensing, the
SZ effect, and numerical simulations, will provide a better
understanding of the regime where hydrostatic equilibrium breaks down
and clusters are still being assembled.

\section*{Acknowledgments}

We thank James Graham, Thomas Reiprich, Steve Allen, Richard
Mushotzky, and Mark Bautz for helpful discussions and Roderick
Johnstone for much assistance. MRG acknowledges a
Herschel Smith Scholarship, HRR is supported by STFC, and ACF thanks
the Royal Society for support. This research has made use of data
obtained from the Suzaku satellite, a collaborative mission between
the space agencies of Japan (JAXA) and the USA (NASA).

\bibliographystyle{mn2e}
\bibliography{pks0745}

\end{document}

%% file: defs.tex
\def\ks{{\rm\thinspace ks}}

\def\gyr{{\rm\thinspace Gyr}}

\def\kev{{\rm\thinspace keV}}

\def\kpc{{\rm\thinspace kpc}}
\def\mpc{{\rm\thinspace Mpc}}

\def\kmps{{\rm\thinspace km \thinspace s^{-1}}}

\def\keV{{\rm\thinspace keV}}

\def\msun{{\rm\thinspace M_{\odot}}}

\def\zsun{{\rm\thinspace Z_{\odot}}}

\def\nh{{N_{\rm H}}}
\def\cmsq{{\rm\thinspace cm^{-3}}}

\def\pownorm{{\rm\thinspace photons\thinspace keV^{-1}\thinspace cm^{-2}\thinspace s^{-1}}}
\def\mknorm{{\rm\thinspace cm^{-5}\thinspace arcmin^{-2}}}

\def\hone{{H\thinspace {\sc i}}}

%% file: pks0745.bbl
\begin{thebibliography}{}

\bibitem[\protect\citeauthoryear{{Afshordi}, {Lin}, {Nagai} \&
  {Sanderson}}{{Afshordi} et~al.}{2007}]{afshordi07}
{Afshordi} N.,  {Lin} Y.-T.,  {Nagai} D.,    {Sanderson} A.~J.~R.,  2007,
  \mnras, 378, 293

\bibitem[\protect\citeauthoryear{{Allen}, {Fabian} \& {Kneib}}{{Allen}
  et~al.}{1996}]{allen96}
{Allen} S.~W.,  {Fabian} A.~C.,    {Kneib} J.~P.,  1996, \mnras, 279, 615

\bibitem[\protect\citeauthoryear{{Allen}, {Rapetti}, {Schmidt}, {Ebeling},
  {Morris} \& {Fabian}}{{Allen} et~al.}{2008}]{allen08}
{Allen} S.~W.,  {Rapetti} D.~A.,  {Schmidt} R.~W.,  {Ebeling} H.,  {Morris}
  R.~G.,    {Fabian} A.~C.,  2008, \mnras, 383, 879

\bibitem[\protect\citeauthoryear{{Allen}, {Schmidt}, {Fabian} \&
  {Ebeling}}{{Allen} et~al.}{2003}]{allen03}
{Allen} S.~W.,  {Schmidt} R.~W.,  {Fabian} A.~C.,    {Ebeling} H.,  2003,
  \mnras, 342, 287

\bibitem[\protect\citeauthoryear{{Anders} \& {Grevesse}}{{Anders} \&
  {Grevesse}}{1989}]{anders89}
{Anders} E.,  {Grevesse} N.,  1989, \gca, 53, 197

\bibitem[\protect\citeauthoryear{{Arnaud}, {Pointecouteau} \& {Pratt}}{{Arnaud}
  et~al.}{2005}]{arnaud05}
{Arnaud} M.,  {Pointecouteau} E.,    {Pratt} G.~W.,  2005, \aap, 441, 893

\bibitem[\protect\citeauthoryear{{Carrera}, {Fabian} \& {Barcons}}{{Carrera}
  et~al.}{1997}]{carrera97}
{Carrera} F.~J.,  {Fabian} A.~C.,    {Barcons} X.,  1997, \mnras, 285, 820

\bibitem[\protect\citeauthoryear{{Cash}}{{Cash}}{1979}]{cash79}
{Cash} W.,  1979, \apj, 228, 939

\bibitem[\protect\citeauthoryear{{Chen}, {Ikebe} \& {B{\"o}hringer}}{{Chen}
  et~al.}{2003}]{chen03}
{Chen} Y.,  {Ikebe} Y.,    {B{\"o}hringer} H.,  2003, \aap, 407, 41

\bibitem[\protect\citeauthoryear{{Comerford} \& {Natarajan}}{{Comerford} \&
  {Natarajan}}{2007}]{comerford07}
{Comerford} J.~M.,  {Natarajan} P.,  2007, \mnras, 379, 190

\bibitem[\protect\citeauthoryear{{De Grandi} \& {Molendi}}{{De Grandi} \&
  {Molendi}}{2002}]{degrandi02}
{De Grandi} S.,  {Molendi} S.,  2002, \apj, 567, 163

\bibitem[\protect\citeauthoryear{{Edge}, {Stewart}, {Fabian} \&
  {Arnaud}}{{Edge} et~al.}{1990}]{edge90}
{Edge} A.~C.,  {Stewart} G.~C.,  {Fabian} A.~C.,    {Arnaud} K.~A.,  1990,
  \mnras, 245, 559

\bibitem[\protect\citeauthoryear{{Eke}, {Cole} \& {Frenk}}{{Eke}
  et~al.}{1996}]{eke96}
{Eke} V.~R.,  {Cole} S.,    {Frenk} C.~S.,  1996, \mnras, 282, 263

\bibitem[\protect\citeauthoryear{{Eke}, {Navarro} \& {Frenk}}{{Eke}
  et~al.}{1998}]{eke98}
{Eke} V.~R.,  {Navarro} J.~F.,    {Frenk} C.~S.,  1998, \apj, 503, 569

\bibitem[\protect\citeauthoryear{{Evrard}, {Metzler} \& {Navarro}}{{Evrard}
  et~al.}{1996}]{evrard96}
{Evrard} A.~E.,  {Metzler} C.~A.,    {Navarro} J.~F.,  1996, \apj, 469, 494

\bibitem[\protect\citeauthoryear{{Fabian} et~al.,}{{Fabian}
  et~al.}{1985}]{fabian85}
{Fabian} A.~C.,  et~al., 1985, \mnras, 216, 923

\bibitem[\protect\citeauthoryear{{Frenk} et~al.,}{{Frenk}
  et~al.}{1999}]{frenk99}
{Frenk} C.~S.,  et~al., 1999, \apj, 525, 554

\bibitem[\protect\citeauthoryear{{Fujimoto} et~al.,}{{Fujimoto}
  et~al.}{2007}]{fujimoto07}
{Fujimoto} R.,  et~al., 2007, \pasj, 59, 133

\bibitem[\protect\citeauthoryear{{Irwin}, {Bregman} \& {Evrard}}{{Irwin}
  et~al.}{1999}]{irwin99}
{Irwin} J.~A.,  {Bregman} J.~N.,    {Evrard} A.~E.,  1999, \apj, 519, 518

\bibitem[\protect\citeauthoryear{{Ishisaki} et~al.,}{{Ishisaki}
  et~al.}{2007}]{ishisaki07}
{Ishisaki} Y.,  et~al., 2007, \pasj, 59, 113

\bibitem[\protect\citeauthoryear{{Kaiser}}{{Kaiser}}{1986}]{kaiser86}
{Kaiser} N.,  1986, \mnras, 222, 323

\bibitem[\protect\citeauthoryear{{Kalberla}, {Burton}, {Hartmann}, {Arnal},
  {Bajaja}, {Morras} \& {P{\"o}ppel}}{{Kalberla} et~al.}{2005}]{kalberla05}
{Kalberla} P.~M.~W.,  {Burton} W.~B.,  {Hartmann} D.,  {Arnal} E.~M.,  {Bajaja}
  E.,  {Morras} R.,    {P{\"o}ppel} W.~G.~L.,  2005, \aap, 440, 775

\bibitem[\protect\citeauthoryear{{Komatsu} et~al.,}{{Komatsu}
  et~al.}{2008}]{komatsu08}
{Komatsu} E.,  et~al., 2008, preprint (arxiv:0803.0547)

\bibitem[\protect\citeauthoryear{{Koyama} et~al.,}{{Koyama}
  et~al.}{2007}]{koyama07}
{Koyama} K.,  et~al., 2007, \pasj, 59, 23

\bibitem[\protect\citeauthoryear{{Liedahl}, {Osterheld} \&
  {Goldstein}}{{Liedahl} et~al.}{1995}]{liedahl95}
{Liedahl} D.~A.,  {Osterheld} A.~L.,    {Goldstein} W.~H.,  1995, \apjl, 438,
  L115

\bibitem[\protect\citeauthoryear{{Loken}, {Norman}, {Nelson}, {Burns}, {Bryan}
  \& {Motl}}{{Loken} et~al.}{2002}]{loken02}
{Loken} C.,  {Norman} M.~L.,  {Nelson} E.,  {Burns} J.,  {Bryan} G.~L.,
  {Motl} P.,  2002, \apj, 579, 571

\bibitem[\protect\citeauthoryear{{Markevitch}, {Forman}, {Sarazin} \&
  {Vikhlinin}}{{Markevitch} et~al.}{1998}]{markevitch98}
{Markevitch} M.,  {Forman} W.~R.,  {Sarazin} C.~L.,    {Vikhlinin} A.,  1998,
  \apj, 503, 77

\bibitem[\protect\citeauthoryear{{McLaughlin}}{{McLaughlin}}{1999}]{mclaughlin%
99}
{McLaughlin} D.~E.,  1999, \aj, 117, 2398

\bibitem[\protect\citeauthoryear{{Mewe}, {Gronenschild} \& {van den
  Oord}}{{Mewe} et~al.}{1985}]{mewe85}
{Mewe} R.,  {Gronenschild} E.~H.~B.~M.,    {van den Oord} G.~H.~J.,  1985,
  \aaps, 62, 197

\bibitem[\protect\citeauthoryear{{Mewe}, {Lemen} \& {van den Oord}}{{Mewe}
  et~al.}{1986}]{mewe86}
{Mewe} R.,  {Lemen} J.~R.,    {van den Oord} G.~H.~J.,  1986, \aaps, 65, 511

\bibitem[\protect\citeauthoryear{{Navarro}, {Frenk} \& {White}}{{Navarro}
  et~al.}{1997}]{navarro97}
{Navarro} J.~F.,  {Frenk} C.~S.,    {White} S.~D.~M.,  1997, \apj, 490, 493

\bibitem[\protect\citeauthoryear{{Neumann}}{{Neumann}}{2005}]{neumann05}
{Neumann} D.~M.,  2005, \aap, 439, 465

\bibitem[\protect\citeauthoryear{{O'Dea} et~al.,}{{O'Dea}
  et~al.}{2008}]{odea08}
{O'Dea} C.~P.,  et~al., 2008, \apj, 681, 1035

\bibitem[\protect\citeauthoryear{{Ostriker}, {Bode} \& {Babul}}{{Ostriker}
  et~al.}{2005}]{ostriker05}
{Ostriker} J.~P.,  {Bode} P.,    {Babul} A.,  2005, \apj, 634, 964

\bibitem[\protect\citeauthoryear{{Pointecouteau}, {Arnaud} \&
  {Pratt}}{{Pointecouteau} et~al.}{2005}]{pointecouteau05}
{Pointecouteau} E.,  {Arnaud} M.,    {Pratt} G.~W.,  2005, \aap, 435, 1

\bibitem[\protect\citeauthoryear{{Ponman}, {Sanderson} \&
  {Finoguenov}}{{Ponman} et~al.}{2003}]{ponman03}
{Ponman} T.~J.,  {Sanderson} A.~J.~R.,    {Finoguenov} A.,  2003, \mnras, 343,
  331

\bibitem[\protect\citeauthoryear{{Pratt}, {B{\"o}hringer}, {Croston}, {Arnaud},
  {Borgani}, {Finoguenov} \& {Temple}}{{Pratt} et~al.}{2007}]{pratt07}
{Pratt} G.~W.,  {B{\"o}hringer} H.,  {Croston} J.~H.,  {Arnaud} M.,  {Borgani}
  S.,  {Finoguenov} A.,    {Temple} R.~F.,  2007, \aap, 461, 71

\bibitem[\protect\citeauthoryear{{Reiprich} \& {B{\"o}hringer}}{{Reiprich} \&
  {B{\"o}hringer}}{2002}]{reiprich02}
{Reiprich} T.~H.,  {B{\"o}hringer} H.,  2002, \apj, 567, 716

\bibitem[\protect\citeauthoryear{Reiprich et~al.,}{Reiprich
  et~al.}{2008}]{reiprich08}
Reiprich T.~H.,  et~al., 2008, preprint (arxiv:0806.2920)

\bibitem[\protect\citeauthoryear{{Roncarelli}, {Ettori}, {Dolag}, {Moscardini},
  {Borgani} \& {Murante}}{{Roncarelli} et~al.}{2006}]{roncarelli06}
{Roncarelli} M.,  {Ettori} S.,  {Dolag} K.,  {Moscardini} L.,  {Borgani} S.,
  {Murante} G.,  2006, \mnras, 373, 1339

\bibitem[\protect\citeauthoryear{{Sanders} \& {Fabian}}{{Sanders} \&
  {Fabian}}{2007}]{sanders07}
{Sanders} J.~S.,  {Fabian} A.~C.,  2007, \mnras, 381, 1381

\bibitem[\protect\citeauthoryear{{Sato} et~al.,}{{Sato}  et~al.}{2007}]{sato07}
{Sato} K.,  et~al., 2007, \pasj, 59, 299

\bibitem[\protect\citeauthoryear{{Schmidt} \& {Allen}}{{Schmidt} \&
  {Allen}}{2007}]{schmidt07}
{Schmidt} R.~W.,  {Allen} S.~W.,  2007, \mnras, 379, 209

\bibitem[\protect\citeauthoryear{{Snowden} et~al.,}{{Snowden}
  et~al.}{1997}]{snowden97}
{Snowden} S.~L.,  et~al., 1997, \apj, 485, 125

\bibitem[\protect\citeauthoryear{{Snowden}, {Mushotzky}, {Kuntz} \&
  {Davis}}{{Snowden} et~al.}{2008}]{snowden08}
{Snowden} S.~L.,  {Mushotzky} R.~F.,  {Kuntz} K.~D.,    {Davis} D.~S.,  2008,
  \aap, 478, 615

\bibitem[\protect\citeauthoryear{{Solovyeva}, {Anokhin}, {Sauvageot},
  {Teyssier} \& {Neumann}}{{Solovyeva} et~al.}{2007}]{solovyeva07}
{Solovyeva} L.,  {Anokhin} S.,  {Sauvageot} J.~L.,  {Teyssier} R.,    {Neumann}
  D.,  2007, \aap, 476, 63

\bibitem[\protect\citeauthoryear{{Suto}, {Sasaki} \& {Makino}}{{Suto}
  et~al.}{1998}]{suto98}
{Suto} Y.,  {Sasaki} S.,    {Makino} N.,  1998, \apj, 509, 544

\bibitem[\protect\citeauthoryear{{Tawa} et~al.,}{{Tawa}  et~al.}{2008}]{tawa08}
{Tawa} N.,  et~al., 2008, \pasj, 60, 11

\bibitem[\protect\citeauthoryear{{Tozzi} \& {Norman}}{{Tozzi} \&
  {Norman}}{2001}]{tozzi01}
{Tozzi} P.,  {Norman} C.,  2001, \apj, 546, 63

\bibitem[\protect\citeauthoryear{{Vikhlinin}, {Kravtsov}, {Forman}, {Jones},
  {Markevitch}, {Murray} \& {Van Speybroeck}}{{Vikhlinin}
  et~al.}{2006}]{vikhlinin06}
{Vikhlinin} A.,  {Kravtsov} A.,  {Forman} W.,  {Jones} C.,  {Markevitch} M.,
  {Murray} S.~S.,    {Van Speybroeck} L.,  2006, \apj, 640, 691

\bibitem[\protect\citeauthoryear{{Vikhlinin}, {Markevitch}, {Murray}, {Jones},
  {Forman} \& {Van Speybroeck}}{{Vikhlinin} et~al.}{2005}]{vikhlinin05}
{Vikhlinin} A.,  {Markevitch} M.,  {Murray} S.~S.,  {Jones} C.,  {Forman} W.,
   {Van Speybroeck} L.,  2005, \apj, 628, 655

\bibitem[\protect\citeauthoryear{{Voigt} \& {Fabian}}{{Voigt} \&
  {Fabian}}{2006}]{voigt06}
{Voigt} L.~M.,  {Fabian} A.~C.,  2006, \mnras, 368, 518

\bibitem[\protect\citeauthoryear{{Voit}}{{Voit}}{2005}]{voit05}
{Voit} G.~M.,  2005, Rev. Mod. Phys., 77, 207

\bibitem[\protect\citeauthoryear{{Voit}, {Kay} \& {Bryan}}{{Voit}
  et~al.}{2005}]{voit05entropy}
{Voit} G.~M.,  {Kay} S.~T.,    {Bryan} G.~L.,  2005, \mnras, 364, 909

\bibitem[\protect\citeauthoryear{{White}, {Navarro}, {Evrard} \&
  {Frenk}}{{White} et~al.}{1993}]{white93}
{White} S.~D.~M.,  {Navarro} J.~F.,  {Evrard} A.~E.,    {Frenk} C.~S.,  1993,
  \nat, 366, 429

\bibitem[\protect\citeauthoryear{{Zhang}, {Finoguenov}, {B{\"o}hringer},
  {Ikebe}, {Matsushita} \& {Schuecker}}{{Zhang} et~al.}{2004}]{zhang04}
{Zhang} Y.-Y.,  {Finoguenov} A.,  {B{\"o}hringer} H.,  {Ikebe} Y.,
  {Matsushita} K.,    {Schuecker} P.,  2004, \aap, 413, 49

\end{thebibliography}
